\documentstyle[aps,prl]{revtex}
\begin{document}
\draft
\title{Shell Model Study of the Double Beta Decays of $^{76}$Ge,
       $^{82}$Se and $^{136}$Xe.}
\author{{\bf E. Caurier$^{*}$, F. Nowacki$^{*}$, A. Poves$^{**}$
         and J. Retamosa$^{**}$}\\
      CRN IN2P3-CNRS/Universit\'e Louis Pasteur BP20\\
      F-67037 Strasbourg-Cedex, France.\\
      $^{**}$ Departamento de F\'{\i}sica Te\'orica C-XI\\
      Universidad Aut\'onoma de Madrid\\
      E-28049 Madrid, Spain.}
\date{\today}
\begin{center}
\maketitle
\begin{abstract}
The lifetimes for the double beta decays of $^{76}$Ge, $^{82}$Se and
$^{136}$Xe are calculated using very large shell model spaces.
The two neutrino  matrix elements obtained are in good agreement with the
present experimental data. For $<m_{\nu}><1$ eV we predict the
following  upper bounds to the half-lives for the neutrinoless mode: 
$T^{(0\nu)}_{1/2}(Ge) > 1.85\,10^{25} yr.$, $T^{(0\nu)}_{1/2}(Se) > 
2.36\,10^{24} yr.$ and $T^{(0\nu)}_{1/2}(Xe) > 1.21\,10^{25} yr$.
These results are the first from a new generation of Shell
Model calculations  reaching O(10$^{8}$) dimensions.
\end{abstract}
\end{center}

The importance of the nuclear double beta decay ($\beta\beta$) is well
established. The two neutrino mode ($\beta\beta_{(2\nu)}$)
that has been proven to be very sensitive to the nuclear correlations,
provides a severe test of the nuclear wave functions. The
neutrinoless mode ($\beta\beta_{(0\nu)}$) is one of the best probes for
the physics beyond the Standard Model. It is particularly suitable to
explore the intrinsic properties of the neutrino like its mass and the
existence of right-handed weak currents
\cite{Hax:84,Doi:85,Boe:92,Moe:94}.

Reasonable values of these lepton number violating parameters can be
extracted  from the experiment provided that accurate nuclear matrix
elements are used. Previous to this work, large Shell Model (SM)
calculations of the $\beta\beta$ decay were only possible in
$^{48}Ca$\cite{Hax:84,Cau:90,Ret:95}. Then, several approximations had
to be used in order to study heavier nuclei.

A weak coupling
approximation was used by Haxton, Stephenson and Strottman in order to
calculate the decays of $^{76}$Ge, $^{82}$Se and $^{128,130}$Te \cite{Hax:84}.
They also invoked the closure approximation to circumvent the calculation of
the 1$^{+}$ states in the intermediate nuclei. Therefore, it is
very difficult to estimate the uncertainty in their results because there
is not an unique method to choose the energy denominator of the two neutrino
matrix element. Using an statistical method to determine
the energy denominators and, setting $g_A = 1$, they
gave reasonable values for the lifetimes in all the cases, except for the
Tellurium isotopes.

The quasiparticle random-phase approximation (QRPA) has also been used
to study the decay of medium and heavy nuclei\cite{Vo:86}.
It was found\cite{Ci:87,Mu:88} that the matrix elements of the Gamow-Teller
and double Gamow-Teller operators were very sensitive to the  particle-particle
interaction. The strength of this interaction (g$_{pp}$) is  treated as a
parameter and fitted to the available $\beta_{\pm}$  data. The introduction
of g$_{pp}$ as a phenomenological factor makes it possible to reproduce
the observed half-lives.
However, the predictivity of the QRPA approach is severely limited because
of the large variation of the relevant matrix element $M^{(2\nu)}_{GT}$
in the physical window for g$_{pp}$.
The neutrinoless mode is less sensitive to the correlations, but there
still exists a factor of three among several predictions for $M^{(0\nu)}_{GT}$
\cite{To:87,Mu:89,St:90}

Very recently the two neutrino mode has been studied in the Shell Model
Monte Carlo method (SMMC)\cite{Rad:95}. For the decay $^{48}Ca\rightarrow
^{48}Ti$ the matrix element is $0.15\pm0.07$ MeV$^{-1}$.
Using the same interaction, the exact diagonalization gives $0.08$ MeV$^{-1}$.
For $^{76}Ge$ a matrix element  $0.13\pm0.05$ MeV$^{-1}$ is found.

The aim of this paper is  the study of the $\beta\beta$ decay of
$^{76}$Ge, $^{82}$Se and $^{136}$Xe in the SM framework.
This study presents a new generation of SM calculations that have recently become
feasible due to the availability of a very performing SM code\cite{Cau:89}
that makes it possible to carry the calculations in an ordinary workstation.
Similar importance has the development of a minimal phenomenological
treatment of the realistic interactions in order to cure their bad saturation
properties \cite{Zuk:95}. We believe that these calculations
represent a real improvement over the old SM and QRPA calculations

For $^{76}$Ge and $^{82}$Se the Schrodinger equation is solved in a
valence space that consists of the following orbitals: $p_{3/2}$,
$f_{5/2}$,  $p_{1/2}$ and $g_{9/2}$. $^{136}Xe$ is studied in
a valence space that is made of the $d_{5/2}$, $s_{1/2}$, $g_{7/2}$,$d_{3/2}$
and $h_{11/2}$ shells. In table \ref{tab1} we show the
number of active particles as well as the dimensions reached in each
case. In the first space t measures the number of particles that are
allowed to  jump from the  pf subshells to the g$_{9/2}$ orbital. In the
case of  A=136 it represents the number of particles that can be
excited from the lower $g_{7/2}$ and $d_{5/2}$ subshells to the
remaining three orbitals. In both cases "full" indicates that
no limitations are imposed. Later on we shall analyse the convergence
of the relevant matrix elements from t=0 to the full calculation, except
in A=76 where we are limited to t=4(Ge and Se) and  t=5(As) because
the full  calculation is out - but not far - of reach.

We use as starting point a G-matrix from Kuo\cite{Dea} to calculate the
wave functions of the A=76,82 nuclei . To fix the interaction the monopole
parameters are fitted to the energy levels of the Ni isotopes and the
N=50 isotones\cite{Now:95}. The main monopole changes amount to weaken the
interaction among the pf orbits and the $g_{9/2}$ shell.
For $^{136}Xe$, the interaction is the G-matrix obtained from the
the Bonn potential\cite{Bonn} with monopole modifications in order to
reproduce the spectroscopy of the N=82 isotones\cite{Now:95}.
The secular problem is solved using the Lanczos algorithm.

The eigenstates are used for calculating the matrix elements of the
weak interaction inducing the $\beta\beta$ decays.
We describe the weak processes in the nucleus by an effective hamiltonian
proposed by Doi {\em et al.}\cite{Doi:85}, that consists of V and A currents
and that is compatible with $SU(2)_L \otimes SU(2)_R \otimes U(1)$ grand
unification models.
\begin{equation}
H_w = \frac{G}{\sqrt{2}}[j_{L}^{\mu} (J_{L\mu}+\chi J_{R\mu})^{+}
+ j_{R}^{\mu} (\eta J_{L\mu} + \lambda J_{R\mu})^{+} ] + h.c. ,
\end{equation}
In addition to the right-left and right-right coupling constants, the
neutrino mass (implicit in the leptonic currents) also breaks the
maximal parity violation of the standard theory. This hamiltonian
led us to an scenario where  there is no maximal parity violation
and where the leptonic number is not preserved. The neutrinoless double
beta decay is mediated by (virtual) massive Majorana neutrinos. Moreover
the description of the two neutrino mode is essentially the same as that
of the Standard Model. This is so because the relevant contributions
come from the standard left-left coupling.
With regard to the hadronic currents, all the terms up to order v/c
are included because, as first noticed by Tomoda {\em et al.}\cite{To:86},
they contribute significantly to the zero neutrino mode.

{\em $(\beta\beta)_{2\nu}$ results}: The half-life can be
approximated as
\begin{equation}
[T_{\frac{1}{2}}^{(2\nu)}(0^{+}->0^{+})]^{-1} = G \left|M^{(2\nu)}_{GT}
\right|^2,
\end{equation}
where G is a Integral kinematical factor and $M^{(2\nu)}_{GT}$
is the usual energy weighted double Gamow-Teller matrix element.
We use the bare Gamow-Teller operator in the definition of the matrix element
and use $g_A=1$  in the integral kinematic factor.
The exact definition of this matrix element as well
as the algorithm that we use to calculate it can be found in reference
\cite{Cau:90}.
By means of this algorithm we can get a reliable approximation to the
Gamow-Teller strength in the intermediate nucleus. However, due to the fact
that our model spaces do not include all the spin partner orbits we can not
exhaust the 3(N-Z) sum rule. In fact, we find:
 $S^{(Ge)}_{-} = 17.14$, $S^{(Se)}_{-} = 21.66$ and $S^{(Xe)}_{-} = 52.30$
compared to the 3(N-Z) values 36, 42 and 84 respectively.
However, as the remaining strength will  appear at high energy its influence
in the description of the two neutrino mode should be small.

We summarize our results in table \ref{tab2}.
This is the only mode whose existence  has been clearly confirmed \cite{Moe:94}
and the lifetimes have been measured for several emitters.
There are two sets of matrix elements. In the first one the
theoretical energies of the 1$^{+}$ states are used whilst, in the
second the spectrum of these states is globally shifted in order to
place the first 1$^{+}$ state at its experimental energy. As it can be
seen in table \ref{tab2} the matrix elements increase a 20$\%$ when
the experimental energies are used. The agreement with the
experimental data  is, in both cases, reasonably good.
Further refinements of the interaction are possible but, we do not
expect major modifications of the matrix elements.

It is worth comment that our closure matrix elements are very
different from those of reference \cite{Hax:84}, 0.68 compared to 2.56
($^{76}$Ge) and 0.74 compared to 1.876 ($^{82}$Se). These
discrepancies also led to quite different effective 1$^{+}$ centroids.

In the A=76,82 region the matrix element increases very slowly as the valence
space is enlarged. Consequently, the half life decreases and there is
a factor two between the t=2 and the final predictions.
This behavior is different to that found in the decay of $^{136}$Xe,
where the matrix element is nearly constant.

It is possible to compare the SM result, 0.14 with the SMMC extrapolation,
0.13$\pm$0.05. Both numbers are very close.  Nevertheless, the
comparison of other relevant quantities ($(M^{2\nu}_{GT})_c$, $S_{+}$,
$\overline{E}$) is not as good. Any further discussion must be delayed
because,  although the valence spaces are identical, the effective
interactions are different.

{\em $(\beta\beta)_{0\nu}$ results}:
In the closure approximation the half-life of the $0^{+}->0^{+}$ decay
can be written as
\begin{eqnarray}
[T_{\frac{1}{2}}^{(0\nu)}(0^{+}->0^{+}]^{-1}&=&\left|M_{GT}^{(0\nu)}
\right|^{2}\{C_{m m}\left(\frac{<m_{\nu}>}{m_{e}}\right)^{2}+
C_{\lambda \lambda}<\lambda>^{2}+ C_{\eta \eta} <\eta>^{2}+\nonumber \\
&+&C_{\lambda m}<\lambda>\frac{<m_{\nu}>}{m_{e}}cos\psi_{1}
+C_{\eta m}<\eta>\frac{<m_{\nu}>}{m_{e}} cos\psi_{2} + \nonumber \\
&+&C_{\lambda \eta} <\lambda><\eta> cos(\psi_{1}-\psi_{2}) \},
\end{eqnarray}
where $<m_{\nu}>$, $<\lambda>$ and $<\eta>$ are the effective lepton
violating parameters, $\psi_{1(2)}$ are the CP phases and the
$C_{xy}$ coefficients are linear combinations of the nine matrix elements
and nine integral factors. A clear and comprehensive definition of them
all can be found in reference \cite{Doi:85}.

The nine matrix elements shown in table \ref{tab3} are calculated in
the light neutrino approximation (see Doi \cite{Doi:85}) and using
$g_A / g_V = 1.25$.
Since this mode has not been observed yet, we take a half-life of 
$10^{25} yr.$, which is very close to the expected experimental limits\cite{NEM:95}.
This makes it possible  to obtain the upper bounds to the three
lepton violating parameters $<m_{\nu}>$, $<\lambda>$ and $<\eta>$.
The results are compiled in table \ref{tab4}. For $<m_{\nu}> < 1 eV$
we find  the following lower bounds: $T^{(0\nu)}_{1/2}(Ge) > 1.85\,10^{25}
yr.$ $T^{(0\nu)}_{1/2}(Se) > 2.36\,10^{24} yr.$ and $T^{(0\nu)}_{1/2}(Xe) >
1.21\,10^{25} yr$.
Although the result for $^{76}Ge$ has been obtained in a t=4 calculation,
we expect that the matrix element and therefore the half-life are close
to convergence, as it is the case in $^{82}Se$. 
Being the $(\beta\beta)_{2\nu}$ mode a background for the detection of
the neutrinoless mode the ratio $T^{(2\nu)}_{1/2} / T^{(0\nu)}_{1/2}$ is
very important. Notice that, among the three studied nuclei, this
ratio is minimum for $^{82}Se$ and, therefore, it can be considered a good
candidate for the detection of the zero neutrino mode.

There are discrepancies among different calculations of the
$\beta\beta_{(0\nu)}$  mode. Table \ref{tab5} compares the $M_{GT}^{(0\nu)}$ 
values predicted by several authors. We see that our matrix elements
are smaller than the others. The only exception are the new results by
Pantis {\em et al.}\cite{Pan:95}, that include to some extent the
neutron-proton pairing interaction.
Based on the reasonable agreement between theory and experiment for
the spectroscopy of the regions around these $\beta\beta$ emitters and
for the $\beta\beta_{(2\nu)}$ mode, we are also quite confident on our
predictions for the zero neutrino mode.

In summary, we have presented the first large scale SM calculations of the
$\beta\beta$ decays of $^{76}$Ge, $^{82}$Se and $^{136}$Xe. The agreement
with the present experimental data, available for the $\beta\beta_{(2\nu)}$
mode, is reasonable and encouraging. For the $\beta\beta_{(0\nu)}$
mode we have obtained upper bounds to the neutrino mass and to the
coupling constants assuming a lifetime of $10^{25} yr.$, and the lifetimes
predictions for $<m_{\nu}> < 1 eV$.

{\em Acknowledgments}. We thank M. Hjorth-Jensen and T.T. Kuo for
making their G-matrix available to us. We thank D. Dean, P. Vogel
and A. Zuker for their help. This work has been partly supported by the
IN2P3 (France)-CICYT(Spain) agreements and by a grant of DGICYT (Spain)
PB93-263.

\newpage

\begin{table}[top]
\begin{center}
\caption{Dimensions and active nucleons for each nucleus. See text for
the meaning of t.}
\label{tab1}
\vspace*{0.5cm}
\begin{math}\begin{array}{lrrrc}
 &N_{\pi}&N_{\nu}&Dimensions&Truncation\\ \hline
       & &  &       &      \\ [-5pt]
^{76}Ge&4&16&8176629&t=4   \\
^{76}As&5&15&63788368&t=5  \\
^{76}Se&6&14&54625321&t=4  \\
 & & &                      \\
^{82}Se&6&20& 605367& full\\
^{82}Br&7&19&8353667& full \\
^{82}Kr&8&18&70757366&full  \\
 & & &                 \\
^{136}Xe&4&32&1504& full \\
^{136}Cs&5&31&212338&full  \\
^{136}Ba&6&30&13139846&full  \\ \hline
\end{array}\end{math}
\end{center}
\end{table}

\begin{table}[bottom]
\begin{center}
\caption{Two neutrino matrix elements and half lives.
$M^{(2\nu)}_{GT}$ in MeV$^{-1}$ and T$_{1/2}$ in years.}
\label{tab2}
\vspace*{0.5cm}
\begin{math}\begin{array}{lcccccccccc}\hline\hline
Decay&t&\multicolumn{2}{c}{M^{(2\nu)}_{GT}}&(M^{(2\nu)}_{GT})_{exp}&\multicolumn{2}{c}{T^{2\nu}_{1/2}}&(T^{2\nu}_{1/2})_{exp} &(M^{(2\nu)}_{GT})_{c}&S_{-}&S_{+}\\[-5pt]
     & &             &            & &             &            & & & \\[-5pt]
     & &\Delta E(exp)&\Delta E(th)& &\Delta E(exp)&\Delta E(th)& & &\\[5pt]\hline
                           & 0 &0.000& 0.000&   &   &   &   &0.000&16.73&0.000 \\
^{76}Ge\rightarrow^{76}Se  & 2 &0.112&0.088&&5.678\,10^{21}&9.197\,10^{21}&   &0.465& 17.13&0.146\\
                           & 4 &0.180&0.140&   &2.198\,10^{21}&3.634\,10^{21}& &0.676&17.14& 0.258 \\
                           &full&   &   &0.22&   &   &1.80\,10^{21}&
                             &  &  \\[-5pt]
     & &             &            & &             &            & & & \\[-5pt]
                           & 0   &0.000&0.000&   &   & &  &0.000&21.66 & 0.000  \\
^{82}Se\rightarrow^{82}Kr  & 2 &0.128&0.102&  &1.312\,10^{20}&2.065\,10^{20}& & 0.483&21.61&0.121 \\
                           & 4 &0.198&0.155&  &5.482\,10^{19}&8.946\,10^{19}&  & 0.745&21.56&0.209 \\
                           &full&0.208&0.164&0.14&4.968\,10^{19}&7.991\,10^{19}&1.08\,10^{20}&0.799&21.55&0.226\\
     & &             &            & &             &            & & & \\[-5pt]
                           & 0 &0.026&0.028&  &2.487\,10^{21}&2.455\,10^{21}&  &0.106& 52.75&0.004\\
^{136}Xe\rightarrow^{136}Ba& 2 &0.036&0.039&  &1.485\,10^{21}&1.265\,10^{21}&  &0.178& 52.37&0.007 \\
                           & 4 &0.032&0.035&  &1.879\,10^{21}&1.571\,10^{21}&   &0.146& 52.30&0.008\\
                           &full&0.031&0.034&<0.06&2.003\,10^{21}&1.665\,10^{21}& > 5.60\,10^{20}&0.143&52.30&0.008\\[1pt]\hline
\end{array}\end{math}
\end{center}
\end{table}

\clearpage

\begin{table}[here]
\begin{center}
\caption[]{Nuclear matrix elements for the $(\beta\beta)_{(0\nu)}$ mode.
$X_{GT}^{\omega} = 2 - X_{GT}^{'}$, $X_{F}^{\omega} = 2X_{F} - X_{F}^{'}$.}
\label{tab3}
\vspace*{0.5cm}
\begin{math}\begin{array}{llrrrr}\hline\hline
\multicolumn{6}{c}{}\\[-6pt]
M.E.& &t=0\,&t=2\,&t=4\,&Full\,\\[5pt] \hline
\multicolumn{6}{c}{}\\[-6pt]
M_{GT}^{(0\nu)}&^{76}Ge&0.721&1.294&1.568&  \\[2pt]
&^{82}Se&0.505&1.340&1.846&1.970\\[2pt]
&^{136}Xe&0.484&0.630&0.649&0.651\\[2pt]
\chi_{F}&^{76}Ge&-0.068&-0.098&-0.106&  \\[2pt]
&^{82}Se&-0.101&-0.107&-0.107&-0.108\\[2pt]
&^{136}Xe&-0.172&-0.156&-0.157&-0.158\\[2pt]
\chi_{GT}^{'}&^{76}Ge&1.074 &1.107 &1.115 &  \\[2pt]
&^{82}Se&1.069&1.114&1.119&1.120\\[2pt]
&^{136}Xe&1.103&1.106&1.099&1.097\\[2pt]
\chi_{F^{'}}&^{76}Ge&-0.060&-0.102&-0.109&  \\[2pt]
&^{82}Se&-0.104&-0.112&-0.112&-0.112\\[2pt]
&^{136}Xe&-0.184&-0.166&-0.167&-0.167\\[2pt]
\chi_{T}&^{76}Ge&0.186&0.043&0.017&    \\[2pt]
&^{82}Se&0.156&0.049&0.031&0.028\\[2pt]
&^{136}Xe&0.039&-0.006&-0.031&-0.031\\[2pt]
\chi_{P}&^{76}Ge&-1.435&-0.832&-0.544&   \\[2pt]
&^{82}Se&1.710&0.852&0.574&0.494\\[2pt]
&^{136}Xe&0.898&0.411&0.280&0.256\\[2pt]
\chi_{R}&^{76}Ge&0.761&0.707&0.684&    \\[2pt]
&^{82}Se&0.873&0.706&0.683&0.680\\[5pt]
&^{136}Xe&0.780&0.872&0.942&0.955\\[2pt] \hline
\end{array}\end{math}
\end{center}
\end{table}

\begin{table}[top]
\begin{center}
\caption[Upper bounds of the parameters of the weak hamiltonian density]
{Shell Model predictions for the upper bounds to the three lepton violating
parameters (LPV): the effective mass $<m_{\nu}>$ and the coupling constants
$<\lambda>$ and $<\eta>$ for $T_{1/2} > 10^{25}yr.$}
\label{tab4}
\vspace*{0.5cm}
\begin{math}\begin{array}{rlcccc}\hline\hline
\multicolumn{6}{c}{}\\[-6pt]
& &t=0\,&t=2\,&t=4\,\,&Full\,\\[5pt] \hline
\multicolumn{6}{c}{}\\[-6pt]
^{76}Ge&<m_{\nu}>(eV)  &2.98&1.60&1.32&    \\[2pt]
       &<\eta>10^{8}   &2.78&1.83&1.65&    \\[2pt]
       &<\lambda>10^{6}&4.31&2.65&2.24&     \\[12pt]
^{82}Se&<m_{\nu}>(eV)  &1.93&0.72&0.52&0.49 \\[2pt]
       &<\eta>10^{8}   &3.03&1.22&0.86&0.79\\[2pt]
       &<\lambda>10^{6}&1.93&0.82&0.61&0.57 \\[12pt]
^{136}Xe&<m_{\nu}>(eV) &1.47&1.14&1.10&1.10 \\[2pt]
       &<\eta>10^{8}   &3.10&1.77&1.53&1.49 \\[2pt]
       &<\lambda>10^{6}&1.88&1.53&1.50&1.49 \\[2pt] \hline
\end{array}\end{math}
\end{center}
\end{table}

\begin{table}[bottom]
\begin{center}
\caption[]{SM versus QRPA M$_{GT}^{(0\nu)}$ matrix elements.}
\label{tab5}
\vspace*{0.5cm}
\begin{math}\begin{array}{rccccccc}\hline\hline
\multicolumn{6}{c}{}\\[-6pt]
&This\,\, Work&Haxton\cite{Hax:84}&Tomoda\cite{To:87}&Muto\cite{Mu:89}&Staut\cite{St:90}&Pantis\cite{Pan:95}\\[5pt]\hline
\multicolumn{6}{c}{}\\[-6pt]
^{76}Ge&1.568&4.180&3.355&3.014 &10.910&1.846\\[5pt]
^{82}Se&1.970&3.450&3.055&2.847 &      &1.153\\[5pt]
^{136}Xe&0.651&   & 1.120&     &       &1.346\\ \hline
\end{array}\end{math}
\end{center}
\end{table}

\end{document}